\documentclass[11pt,notitlepage]{article}

\usepackage[T1]{fontenc}
\usepackage{lmodern}
\usepackage{microtype}

\usepackage{amsmath,amssymb}

\usepackage{graphicx}
\graphicspath{{}}

\usepackage[margin=2.7cm]{geometry}
\usepackage{setspace}
\usepackage{enumitem}
\usepackage[hidelinks]{hyperref}

\makeatletter
\renewcommand{\maketitle}{%
  \begingroup
  \renewcommand{\thefootnote}{\fnsymbol{footnote}}%
  \parindent=0pt
  \begin{center}
    {\LARGE\bfseries \@title\par}
    \vskip 0.8em
    {\normalsize \@author\par}
    \vskip 0.6em
    {\normalsize \@date\par}
  \end{center}
  \endgroup
  \vskip 1.0em
}
\makeatother

\title{Particle Dobble – Exploring the Particle Zoo}
\author{
    Lukas Mientus\textsuperscript{1}, 
    Anna Ruechel\textsuperscript{2}, 
    Karsten Kalke\textsuperscript{2}, \& 
    Andreas Borowski\textsuperscript{2}\\
    \small  
    \textsuperscript{1}Otto von Guericke University Magdeburg, Germany
    \textsuperscript{2}University of Potsdam, Germany}
\date{}

\begin{document}

% --- publisher note (kurz halten) ---
{\footnotesize\noindent
This is the accepted manuscript of an article accepted for publication in \emph{The Physics Teacher}.\\
\copyright\ 2026 American Association of Physics Teachers.\\
The final published version will be available at \url{https://doi.org/XXXX}.%
\par}

\vspace{2.5em}

\enlargethispage{\baselineskip}

\maketitle

\vspace{1.5em}

Understanding particle physics can be a daunting challenge for students due to its abstract nature and counterintuitive principles. Concepts such as quantum mechanics, the wave-particle duality, and the Higgs field often conflict with everyday experiences, leading to widespread misconceptions (Styr, 1996; Sigh, 2008; Sigh \& Marshman, 2015). To address these challenges, we have developed Particle Dobble, a science education project aimed at upper high school and early university students that gamifies key aspects of the Standard Model of Particle Physics.

Inspired by the popular card game Dobble (Asmodee GmbH, n.d.), Particle Dobble transforms particle physics into a hands-on, engaging learning experience. The game's cards feature six unique symbols each representing particles and concepts such as quarks, leptons, bosons, and the Higgs mechanism. The game includes small supplementary educational materials, such as particle descriptions and game instructions, freely available under an Open Access license. Particle Dobble represents a novel fusion of education and entertainment, offering a practical tool for educators and a fun, accessible entry point into the fascinating world of fundamental particles for students and enthusiasts alike.

\section{Standard Model of Particle Physics}

The Standard Model of Particle Physics is our best explanation of the tiny building blocks of the universe and how they interact (excluding gravity). It describes two main types of particles: fermions, which make up matter, and bosons, which carry the forces between them.

Fermions include quarks and leptons. Quarks are the components of protons and neutrons, which make up atomic nuclei and whimsically named up, down, strange, charm, top, and bottom. There are six types of quarks, with combinations forming particles like the proton (two up quarks and one down quark) and the neutron (one up and two down quarks). Quarks interact through all known forces and are never found alone; they’re always tightly bound in groups due to the strong force. Leptons include the familiar electron, as well as the heavier muon and tau, each with a partner called a neutrino. Neutrinos are nearly massless and interact very weakly with matter, making them difficult to detect. Bosons are the force carriers. The photon is the boson that carries the electromagnetic force, while the graviton – still only hypothetical boson and not (yet) a particle of the particle zoo – is thought to carry the gravitational force. The W and Z bosons are responsible for the weak force, which is involved in radioactive decay. The gluon is the boson that binds quarks together via the strong force. In 2012, scientists discovered the Higgs boson, thereby proving the existence of the Higgs field. Particles that interact with the Higgs field have mass, while particles that do not interact with it are massless. The Higgs boson is a quantum excitation or a ripple of the Higgs field, like a tiny blip you could observe in an otherwise smooth ocean.

While powerful, the Standard Model is incomplete. It doesn’t include gravity, doesn’t explain dark matter or dark energy, and doesn’t fully account for neutrino masses or why there is more matter than antimatter in the universe. Nevertheless, the Standard Model is a triumph of human understanding and a springboard for future discovery. In educational contexts, however, its abstract nature makes it difficult to present without encouraging classical analogies that lead to misconceptions.

\section{Students Understanding of Particles}

Students often develop misconceptions about particle physics because the subject is abstract and counterintuitive, clashing with their everyday experiences (Styr, 1996). Understanding these misconceptions is crucial for developing effective teaching strategies.

A common misunderstanding is imagining particles in the quantum world like electrons as tiny, solid spheres similar to classical billiard balls. In reality, quantum particles are better described as excitations in quantum fields, showing both particle and wave behavior. Since particles don't have defined positions or paths until measured, students often struggle with concepts like probability distributions and wavefunctions. Another source of confusion is how forces work in particle physics. Students tend to think of forces as pushes and pulls, which makes it hard to grasp the role of force-mediating particles like photons or gluons. The idea that particles interact by exchanging virtual bosons is especially difficult, as virtual particles aren't directly observable and behave unlike anything in the classical world.

The Higgs boson is often misunderstood too. Many believe it "gives mass" to particles directly, as if transferring matter. In fact, particles gain mass by interacting with the Higgs field, which fills all space. Popular media often oversimplifies this, reinforcing the misconception. Misunderstandings about antimatter are also common. Influenced by science fiction, students often see antimatter as rare or dangerously explosive. In reality, antimatter consists of antiparticles – mirror versions of regular particles with opposite charge. It annihilates with matter to release energy, but this is a natural process governed by well-known physics, such as in cosmic rays.

Students also find quarks difficult to conceptualize. They often think quarks cannot be observed alone due to technological limits, rather than understanding color confinement – a fundamental principle of quantum chromodynamics. Color confinement means that quarks are permanently bound together by the strong force, preventing them from existing in isolation. Quarks can only be found within composite particles such as protons and neutrons. The structure of six quark flavors can also seem arbitrary, rather than reflecting deep symmetries in the Standard Model. Finally, the probabilistic nature of quantum mechanics is often misread as randomness or chaos. Students may confuse uncertainty with disorder, not realizing that quantum systems follow strict probabilistic rules grounded in mathematics.

To help students, educators should go beyond equations to build strong conceptual foundations. Visualizations, analogies, and interactive models can connect abstract ideas to students' intuitive thinking. Games such as Particle Dobble can serve as accessible entry points that replace literal visualizations with symbolic ones, encouraging students to abandon the notion of particles as miniature billiard balls. Addressing the misleading notion of particles as simple spheres is a helpful first step – using symbolic and creative representations can support learning, especially given the wide variety of particle types and properties.

\section{Learning with the Game}

Gamification in science teaching is an innovative instructional strategy that integrate elements of games, such as competition and interaction, into educational settings. These approaches aim to make learning more engaging, interactive, and effective by leveraging the motivational power of games to foster deeper understanding and enthusiasm for science (Magney, 1990; Kapp, 2012). These game elements tap into students’ intrinsic and extrinsic motivation, transforming mundane tasks into rewarding experiences (Kettler \& Kauffeld, 2019). Nevertheless, not all learners are equally comfortable with competitive or game-based environments. Some may find them stressful, particularly if they struggle with the subject matter or lack gaming experience. Teachers should implement games in learning settings as additional opportunities for getting in touch with less everyday topics. For this reason, we developed Particle Dobble as an optional learning aid. The primary educational objective is not to teach quantum mechanics directly, but to dispel the common Newtonian misconception that elementary particles behave as tangible, interacting objects. The game addresses one of the key misconceptions within the complex “particle zoo”: the idea that elementary particles are solid spheres. By replacing literal visual representations with playful, abstract symbols (e.g., hearts, shoes, blobs), students experience that “particle identity” does not depend on physical form or classical imagery. Instead, the symbolic representations prompt reflection and discussion, reinforcing that particles are quantum entities best understood through their interactions and properties rather than shapes. (see figure 1).

\begin{figure}[t]
\centering
\includegraphics[width=0.8\textwidth]{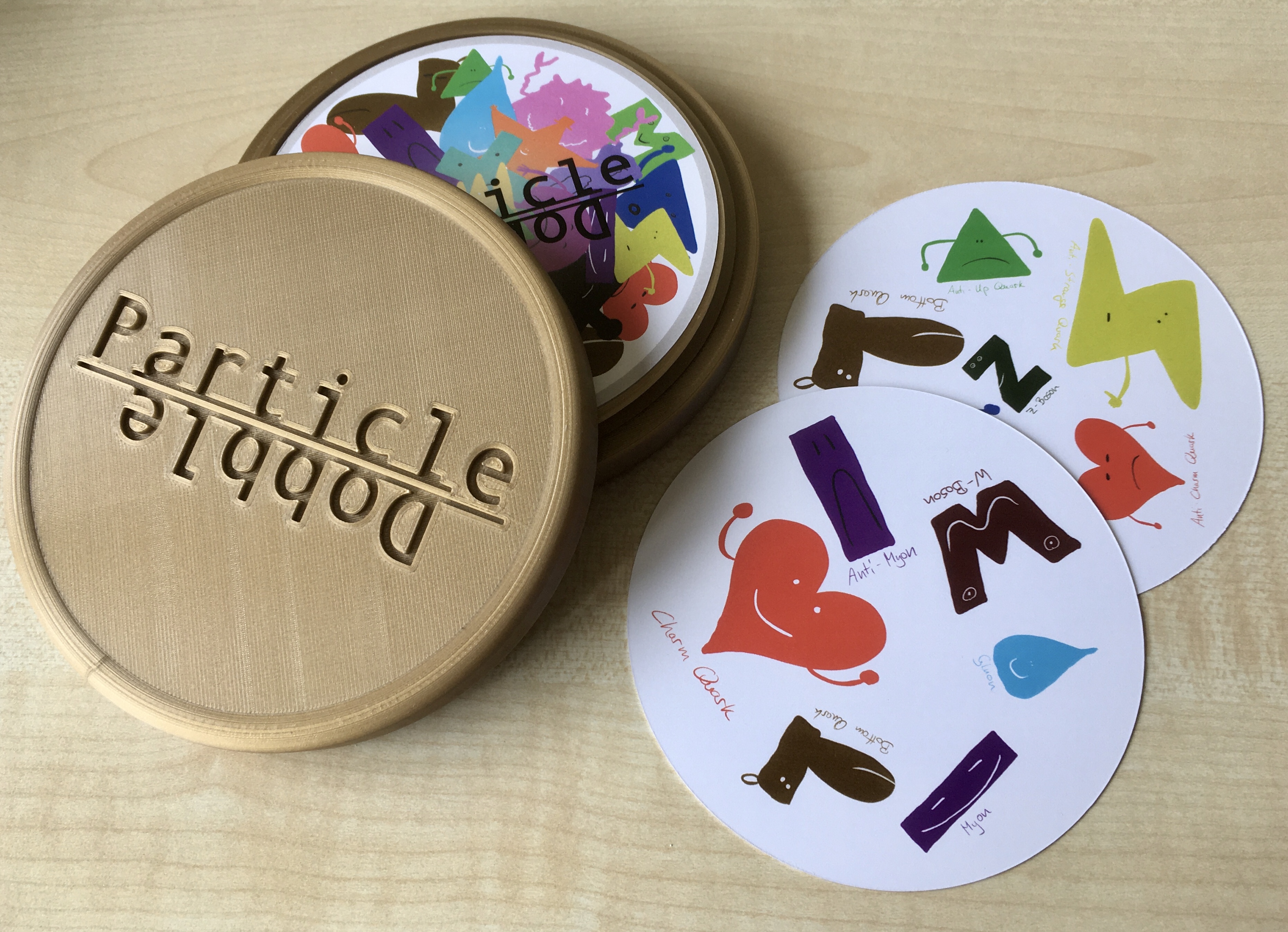}
\caption{Card Deck and Box of Particle Dobble}
\label{fig:sim-map}
\end{figure}

In Particle Dobble, particles are identified purely by their icons. This is deliberate: by using whimsical, non-literal symbols, the game replaces “tiny marbles” with clearly conventional signs. That design choice supports a conceptual shift away from Newtonian thinking without requiring formal quantum mechanics. Each icon-to-particle mapping is arbitrary by design. Students see that an electron or a gluon can be “this symbol” only because we agree it is and not because particles have fixed shapes. That visual constructivism acts as a refutational prompt: if an icon can be anything, then a particle cannot be a tiny object with everyday attributes in the Newtonian sense. Furthermore, the speeded “spot-and-name” mechanic gives repeated retrieval practice on particle names and families (quarks, leptons, and bosons). The game does not simulate quantum dynamics or teach the definition of a particle. Its purpose is earlier in the learning sequence: to unsettle the everyday, Newtonian picture and install a more flexible, non-literal representational stance. With that stance in place, subsequent instruction can introduce ideas such as quark confinement and field excitations without students reverting to little balls.

\section{The Game and the Math}

From a design perspective, Dobble is a fast-paced card game that challenges players' observational skills and quick reflexes. The game's original version consists of a deck of circular cards, each featuring eight symbols of varying sizes. Every card is unique, but each pair of cards shares exactly one symbol in common. The game aims to identify the matching symbol between two cards faster than your opponents. Dobble can be played in several mini-games, each with its own rules, but the core mechanic remains the same: spot the matching symbol and interact as mentioned.  For one version, for example, each person receives a face-down card. The remaining cards form a face-up pile in the middle. Then, all players flip the face-down cards and try to identify the particle shared by the middle card and the hand card. If identified, the player names the particle, take the card from and use it as the new hand card. Whoever has the most cards at the end wins the round. The original symbols on the cards are diverse and whimsical, including items like a snowman, a cat, a lightning bolt, and more. This variety, combined with the varying orientations and sizes of the symbols, makes identifying matches both challenging and entertaining. The game accommodates 2 to 8 players, making it suitable for small or large groups, and its rules are simple enough for children yet engaging enough for adults.

The game Dobble is based on principles of combinatorial design and finite geometry, ensuring that every pair of cards in the deck shares exactly one symbol in common. This mathematical design is rooted in the concept of finite projective planes and their properties. A finite projective plane is a geometric construct in which points and lines are arranged to satisfy specific rules. In the context of Dobble, the "points" are the symbols on the cards, and the "lines" represent the cards themselves. The mathematical framework ensures that any two lines (cards) intersect at exactly one point (symbol), which forms the basis of Dobble’s gameplay. The construction of the Dobble deck typically corresponds to a finite projective plane of order \textit{n}. The order \textit{n} determines the number of symbols and cards in the deck:

\begin{enumerate}
  \item Each card contains \textit{n + 1} symbols.
  \item The total number of symbols in the game is \textit{n\textsuperscript{2} + n + 1}.
  \item The total number of cards in the deck is also \textit{n\textsuperscript{2} + n + 1}.
\end{enumerate}

\noindent
For example, in the Kids edition (as the basis for Particle Dobble), \textit{n = 5}, leading to:

\begin{itemize}
  \item \textit{n + 1 = 6} symbols per card.
  \item \textit{n\textsuperscript{2} + n + 1 = 3} unique symbols and 31 cards.
\end{itemize}

\section{Particle Dobble – Materials}

Based on the Dobble-Kids game variant, we have developed a new version called Particle-Dobble. In this version, each card features six symbols, resulting in 31 unique symbols. Additionally, Particle Dobble includes two special cards, called \textit{Crazies}, that represent dark matter or dark energy. To counteract common misconceptions about particle physics (e.g. that elementary particles are tiny balls or any other object like things from the Newtonian physics), we have designed custom symbols specifically for the particles. These unique representations make the game both educational and engaging. All cards are available for open access under a CC BY-NC-ND license via the link below. For optimal durability and usability, we recommend printing the cards on paper with a weight of at least 300 \textit{g/m\textsuperscript{2}}, as this provides a high-quality tactile experience. We also suggest using a laser printer for printing to avoid smudging, which can occur when inkjet printers are used if the cards are handled frequently. A circle cutter from craft supply stores (\textit{d} = 10.5 cm) is highly recommended for cutting out the circular cards. The downloadable PDF file includes the basic game instructions and a detailed overview of elementary particles. While additional game variants can be easily found online, the included materials serve as a comprehensive starting point. For storing the cards, we recommend a suitable package. To complement the game, we offer a screw-top box design that can be 3D-printed. The corresponding 3D printing file can also be downloaded via a link. We hope you enjoy playing Particle-Dobble and encourage you to share it with others.

\begin{figure}[h]
\centering
\begin{minipage}{0.8\textwidth}
Download all materials via the link below or scan the QR code.\\[0.5em]

\url{https://lukasmientus.github.io/particledobble/material.html}
\end{minipage}
\hfill
\begin{minipage}{0.15 \textwidth}
\centering
\includegraphics[width=\linewidth]{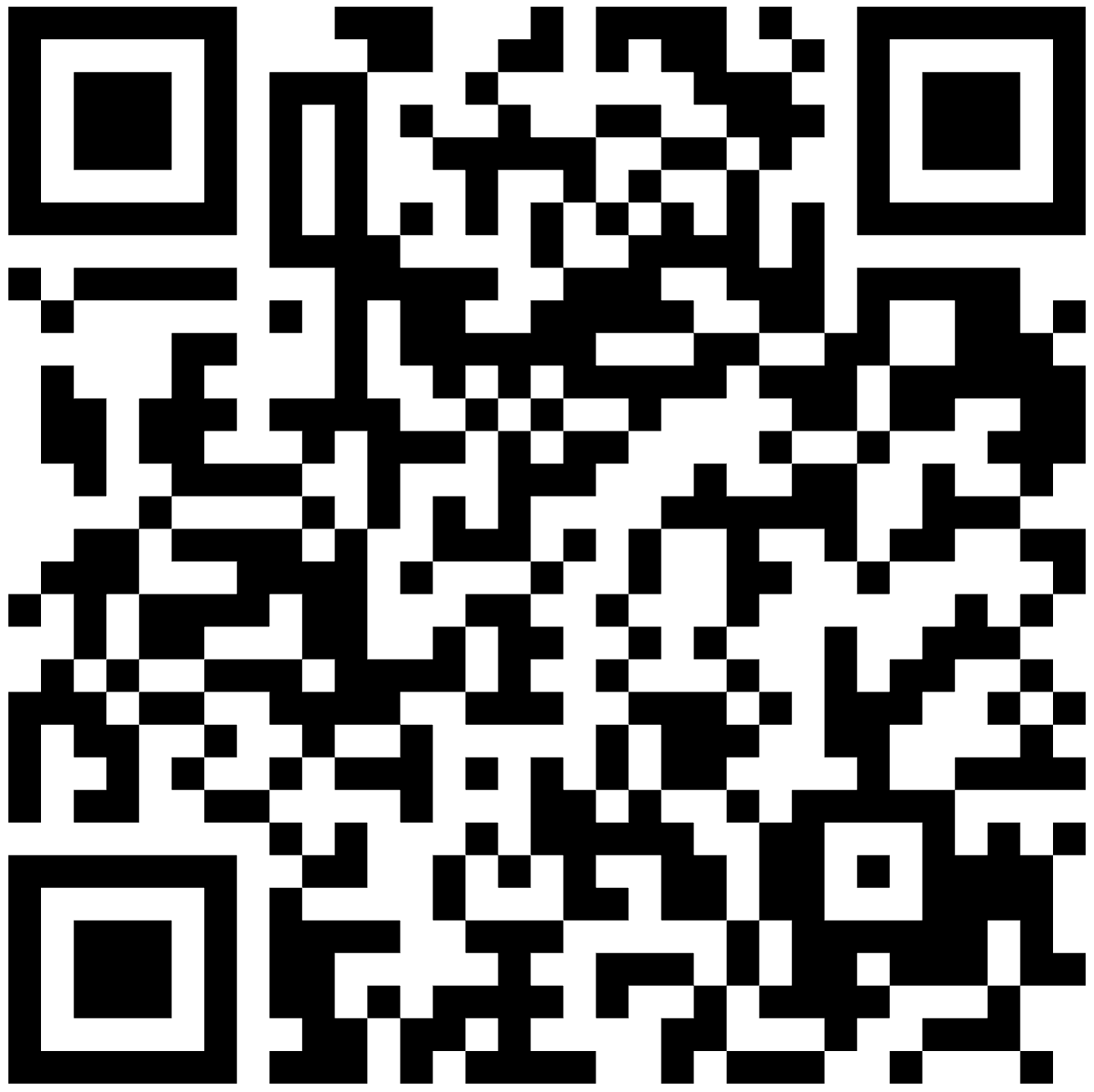}
\end{minipage}
\end{figure}

\section{Conclusion}

Particle Dobble is designed for upper secondary school and introductory undergraduate physics courses, as well as informal learning contexts such as science festivals or outreach events. It assumes a basic familiarity with atomic structure but no formal background in quantum mechanics or particle physics. Educators may use the game as a discussion starter or as a reflective activity following a lesson on fundamental particles. The playful format invites students to let go of overly concrete mental models before they encounter the more formal mathematical framework of quantum mechanics. By combining simple mechanics with abstract representations, Particle Dobble offers a playful yet conceptually grounded way to address one of the most persistent misconceptions in particle physics education. The game encourages students to think relationally rather than pictorially, aligning with the non-classical worldview necessary to grasp modern physics.

% --- References ---
\begingroup
\setstretch{1} % single-spaced references, stable
\sloppy

\fussy
\endgroup
\end{document}